\documentclass[journal]{IEEEtran}

\usepackage[cmex10]{amsmath}
\usepackage{amsfonts}
\usepackage{amssymb}
\usepackage{amsopn}
\usepackage{graphicx}

\newcommand{\sups}[1]{\ensuremath{^{\textrm{#1}}}}
\newcommand{\subs}[1]{\ensuremath{_{\textrm{#1}}}}
\newcommand{\co}[0]{CO\subs{2} }
\newcommand{\bo}[1]{\boldsymbol{#1} }

\begin{document}

\markboth{IEEE Transactions on Instrumentation and Measurement}{Kar}

\title{Accurate Estimation of Gaseous Strength using Transient Data}

\author{Swarnendu~Kar,~\IEEEmembership{Student~Member,~IEEE,}
        and~Pramod~K.~Varshney,~\IEEEmembership{Fellow,~IEEE}
\thanks{The authors are with the Department
of Electrical Engineering and Computer Science, Syracuse University, Syracuse,
NY, 13244 USA, e-mail: \{swkar,varshney\}@syr.edu.}
\thanks{This is a postprint of the original article: S. Kar and P. K. Varshney, ``Accurate Estimation of Gaseous Strength Using Transient Data,'' IEEE Transactions on Instrumentation and Measurement, vol.~60, no.~4, pp.~1197--1205, April 2011 }
\thanks{http://dx.doi.org/10.1109/TIM.2010.2084731}
}

\maketitle

\begin{abstract}
Information about the strength of gas sources in buildings has a number of applications in the area of building automation and control, including temperature and ventilation control, fire detection and security systems. Here, we consider the problem of estimating the strength of a gas source in an enclosure when some of the parameters of the gas transport process are unknown. Traditionally, these problems are either solved by the Maximum-Likelihood (ML) method which is accurate but computationally intense, or by Recursive Least Squares (RLS, also Kalman) filtering which is simpler but less accurate. In this paper, we suggest a different statistical estimation procedure based on the concept of Method of Moments. We outline techniques that make this procedure computationally efficient and amenable for recursive implementation. We provide a comparative analysis of our proposed method based on experimental results as well as Monte-Carlo simulations. When used with the building control systems, these algorithms can estimate the gaseous strength in a room both quickly and accurately, and can potentially provide improved indoor air quality in an efficient manner.
\end{abstract}


%
\IEEEpeerreviewmaketitle

\section{Introduction}
Recent studies have indicated that indoor air quality (IAQ) has a significant effect on the productivity and health of the occupants, e.g., for the U.S., the estimated potential annual savings and productivity gains due to improved indoor air quality are $\$30$ billion to $\$170$ billion \cite{Fis97}. Consequently, a plethora of research is being aimed at improving indoor environmental systems, yet making sure that they are energy efficient and sustainable \cite{Fan00}. One such energy-efficient technique is demand controlled ventilation (DCV). DCV is a well documented method that reduces heating and cooling requirements in buildings by adjusting ventilation rates in response to occupancy \cite{Wang99}, \cite{Sch01}. Determination of occupancy inside a space is essential to implement DCV, and this can be achieved by processing data from various sources like gas sensors, cameras, microphones, or may be even via manual logbooks and electronic calenders. Carbon dioxide (\co) sensors are widely used because of cost considerations and also due to privacy concerns. The basis of using \co for occupancy estimation is established in well-quantified principles of human physiology. All humans, given a similar activity level, exhale \co at a predictable rate based on occupant age and activity level, e.g., \cite{Persily97}, \cite{Ash07}.

The mass-balance model for \co generation in buildings is well established, e.g., \cite{Persily97}, \cite{Aglan03}. An indoor \co measurement provides a dynamic measure of the balance between \co generation in the space, representing occupancy and the amount of low \co concentration in outside air introduced for ventilation (also known as inflow-rate). As a result, the indoor \co increases gradually until it achieves an equilibrium point where the \co produced by people is in balance with the dilution rate. A DCV based control strategy can either act on the transient (pre-equilibrium) data \cite{Fed97} or wait for the steady state to be reached \cite{Lawrence2007} before estimating the occupancy. Since at low inflow-rates it takes longer to reach equilibrium, steady-state algorithms usually involve significant delay \cite{Wang99}. In this paper, we consider using the transient data for occupancy estimation.

The problem of estimating the \co generation rate (or occupancy) and/or inflow parameters using transient data has been considered by several researchers. Most of the available techniques use the differential model of the mass-balance equations for estimation followed by a filtering operation to smooth the estimates. For example, an autoregressive filter was used by \cite{Wang99} and Kalman filters were used by \cite{Fed97} and \cite{Brandes06}. While these techniques are very effective when dealing with dynamic situations, they often produce sub-optimal estimates when the occupancy is constant or slowly varying, as would be the case for meetings, classrooms, etc. As noted in \cite{Fed97}, this deterioration of performance is due to the use of time-derivatives of concentrations. Since the derivatives cannot be measured directly, they must be approximated, and this approximation amplifies high frequency noise.

When the occupancy in a space is constant, the integral model of the mass-balance equations resembles a simple growth curve, whose parameters are related to the occupancy and inflow-rate. Such a model is known to be adequately accurate for several applications \cite{Aglan03}. In this paper, we estimate the occupancy and inflow-rate by estimating the parameters of the growth curve and propose a new approach based on Method of Moments (MME) \cite{Kay93}. Since our MME approach involves the use of time integrals which are more robust to noise, it performs better in terms of estimation accuracy. This fact is demonstrated using both theoretical analysis and experimental data. We also examine the performance of our algorithm in cases when the occupancy is slowly varying or when there is uncertainty regarding the activity levels of the occupants. A preliminary version of this work was presented in \cite{Kar09}.

Though many researchers have suggested procedures to estimate the \co generation rate, performance analysis regarding the uncertainty of estimates has been scarce. One principal goal of this paper is to derive performance bounds that will be useful to practitioners. It is well known in statistical estimation literature (e.g., \cite{Kay93}, \cite{Bar01book}) that maximum likelihood estimator (MLE) has the minimum expected variance and is asymptotically optimal, i.e., attains the Cram\'er-Rao Lower bound (CRLB). In this paper, we derive approximate closed form expressions for the CRLB and also corresponding performance metrics for the KF based procedure (presented in \cite{Fed97}) and the Method of Moments based approach. We demonstrate that for the transient region of operation, the MME estimator is more accurate than the KF counterpart. This improved accuracy can translate to significant reductions in response times and/or reduction in number of sensors.

The rest of the paper is organized as follows. In Section \ref{sec:probform}, we describe the model of the dynamics inside a well-mixed space, formulate our estimation problem, and introduce certain approximations that will aid our theoretical analysis. In Section \ref{sec:relatedwork}, we discuss the MLE and Kalman Filtering based techniques and derive the CRLB and KF performance metrics. In Section \ref{sec:mme}, we describe our time-integral based approach, comment on computational issues and derive the expected variance. In Section \ref{sec:results}, we describe an experimental setup that assumes near ideal conditions and compare the estimation performances. We also present performance comparisons using theoretical results and Monte-Carlo simulations. Concluding remarks are provided in Section \ref{sec:conclude}.

\section{Problem Formulation} \label{sec:probform}

\begin{figure}[htb]
\begin{center}
    \includegraphics[width=9cm]{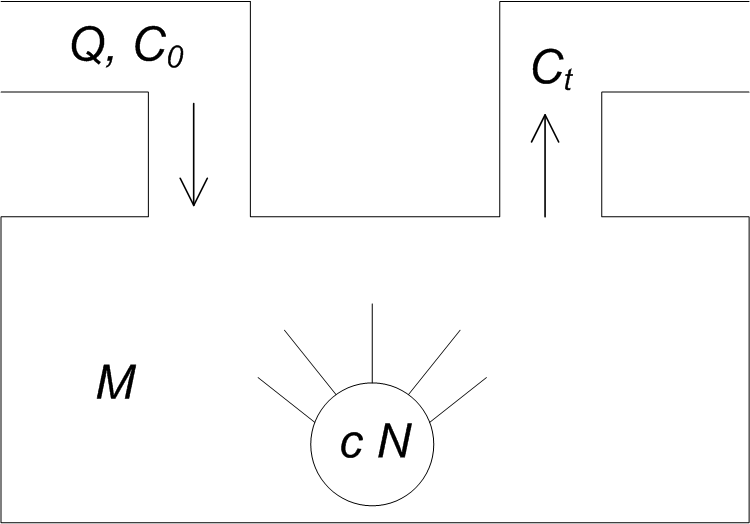}
  \caption{Schematic diagram of an ideal indoor space. $N$ is the number of persons that emit \co at a total rate of $cN$ kg/sec. $C_t,C_0$ are the \co concentration (ppm) of indoor and supply air.}
  \label{fig:schematic}
  \end{center}
\end{figure}

The model of the dynamics inside a well-mixed space is described in this section. This is identical to the well-known single-zone model (e.g., \cite{Aglan03},\cite{Fed97},\cite{Lee01}) and we make the following assumptions - the mass of the air in the space is constant, and the concentration distribution in the space is spatially uniform. Consider the schematic diagram of an indoor space in Figure \ref{fig:schematic}.

Let $N$ denote the number of occupants, $c$ (kg/sec) the \co generation rate per person, $M$ (kg) the mass of air inside the room and $t$ (sec) a particular time instant. Let $Q$ (kg/sec), $C_t$ (ppm or mg/kg) and $C_0$ (ppm) be the airflow rate and \co concentration of indoor and supply air respectively. The number of occupants $N$ and the airflow rate $Q$ are assumed to be unknown quantities that need to be estimated.

A macroscopic mass balance in such a space results in the following mass continuity equation that is widely used by both the academic \cite{Fed97,Lee01} and industrial \cite{Hon07} communities to describe gas transport processes in an indoor environment
\begin{align}
M\dot{C_t}=-Q(C_t-C_0)+cN.
\end{align}
Assuming that there were no gas sources in the room already, the integration yields
\begin{align}
C_t=C_0+(cN/Q) \left(1-\exp (-tQ/M) \right).
\end{align}
Now we reduce the number of unknowns in the model. We note here that $c$ and $M$ are constants and hence can be either known beforehand or obtained via training. We assume $C_0$ to be constant over the interval of interest. Measuring the supply duct concentration, we can obtain a fairly accurate estimate of $C_0$ and we treat it as known. Let the sampling interval be $T_s$ and the total duration of measurements be $T$. This means that the $i$\sups{th} sample is obtained at time $t=i T_s$, where $i=1,2,\ldots n$ and $T = n T_s$. Defining $a_t=C_t-C_0$, our models in differential and integral domains are
\begin{align}
\frac{d a_i}{dt}&= cN-Q a_i, \mbox{ and }  \label{diffmodel} \\
a_i(N,Q)&=(cN/Q)\left(1-\exp (-iT_sQ/M) \right) \label{intmodel},
\end{align}
respectively. We will use the notations $a_i$ and $a_i(N,Q)$ interchangeably throughout this paper. In expressions where the dependence on $N$ and $Q$ needs to be stressed explicitly, we use the longer version of the notation. Next, we characterize the measurement errors. While obtaining the measurement for $C_i$, we assume that the \co sensor is corrupted by a time-independent additive zero-mean Gaussian noise $\epsilon_i$ with variance $\sigma^2$. We assume $\sigma^2$ to be known since the measurement error of sensor noise can be learnt beforehand. Let $y_i$ denote the noise-added observation for  $a_i$, i.e., $y_i=a_i+\epsilon_i$. In vector notations, let $\bo a=[a_1,a_2,\ldots,a_n]'$, $\bo y=[y_1,y_2,\ldots y_n]'$ and $\bo \epsilon=[\epsilon_1,\epsilon_2,\ldots,\epsilon_n]'$, so that
\begin{align}
\bo y=\bo a+\bo \epsilon. \label{def:vecy}
\end{align}
Our goal here is to estimate $\bo \theta := [N,Q]'$, which includes the estimated number of occupants $\widehat{N}$.

\begin{figure}[htb]
\begin{center}
    \includegraphics[width=9cm]{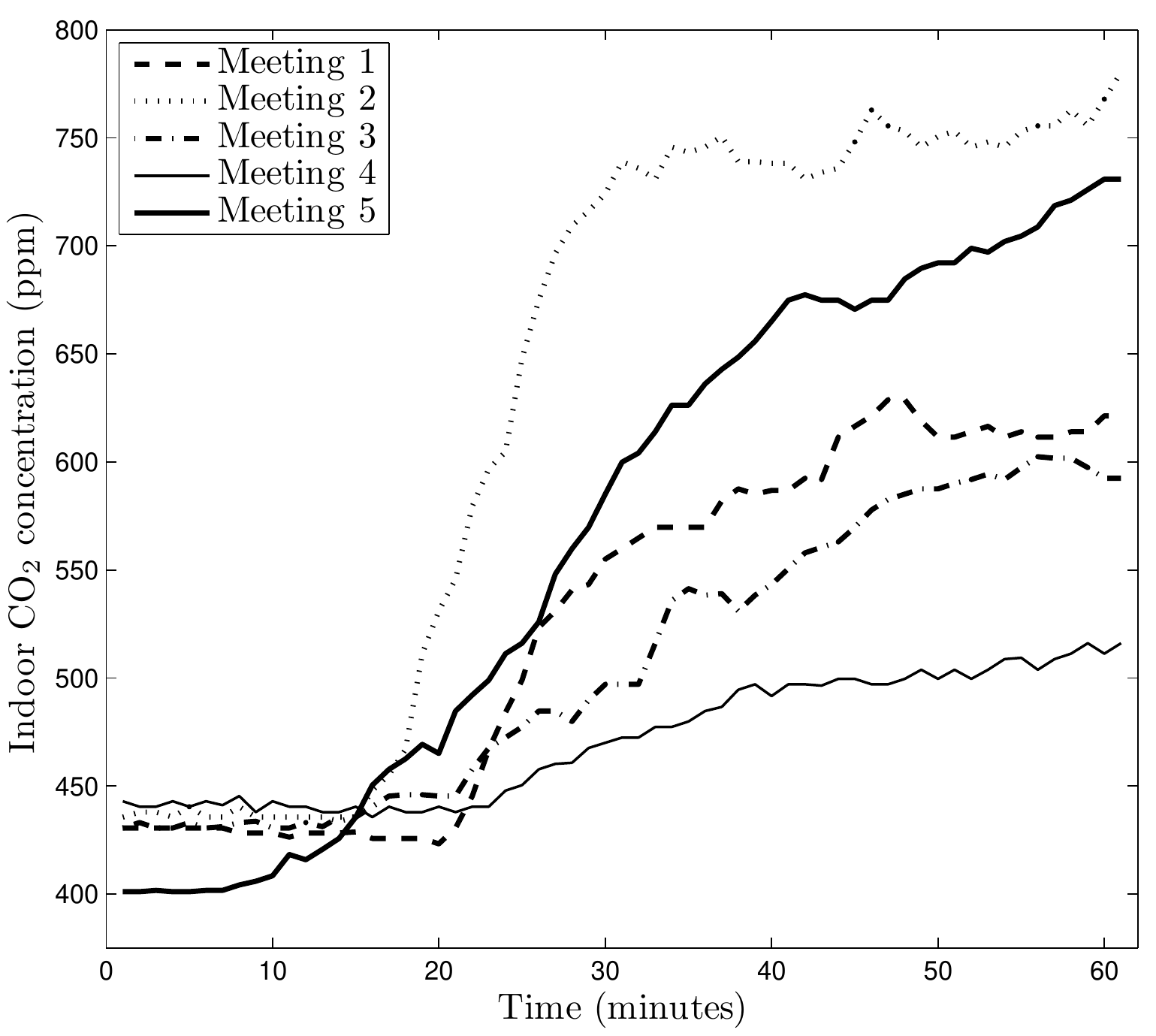}
  \caption{Time-response of the indoor \co concentration during some of the meeting times.}
  \label{fig:utrcco2}
  \end{center}
\end{figure}

The time-series of the indoor \co concentration, as predicted by the mass balance equation stated above, resembles a growth curve that increases gradually before settling down to an equilibrium. For example, the indoor \co concentration during some of the meeting times are shown\footnote{This dataset was acquired during a collaborative research project involving United Technologies Research Center and Syracuse University. Concentration values were sampled every minute from commercially available TI-4GS-22C06 sensors with rated accuracy of $\pm 75$ ppm.} in Figure \ref{fig:utrcco2}. In all cases, there is a region before equilibrium where the measurements increase steadily. We refer to this part as the transient region.

\begin{figure}[htb]
\begin{center}
    \includegraphics[width=9cm]{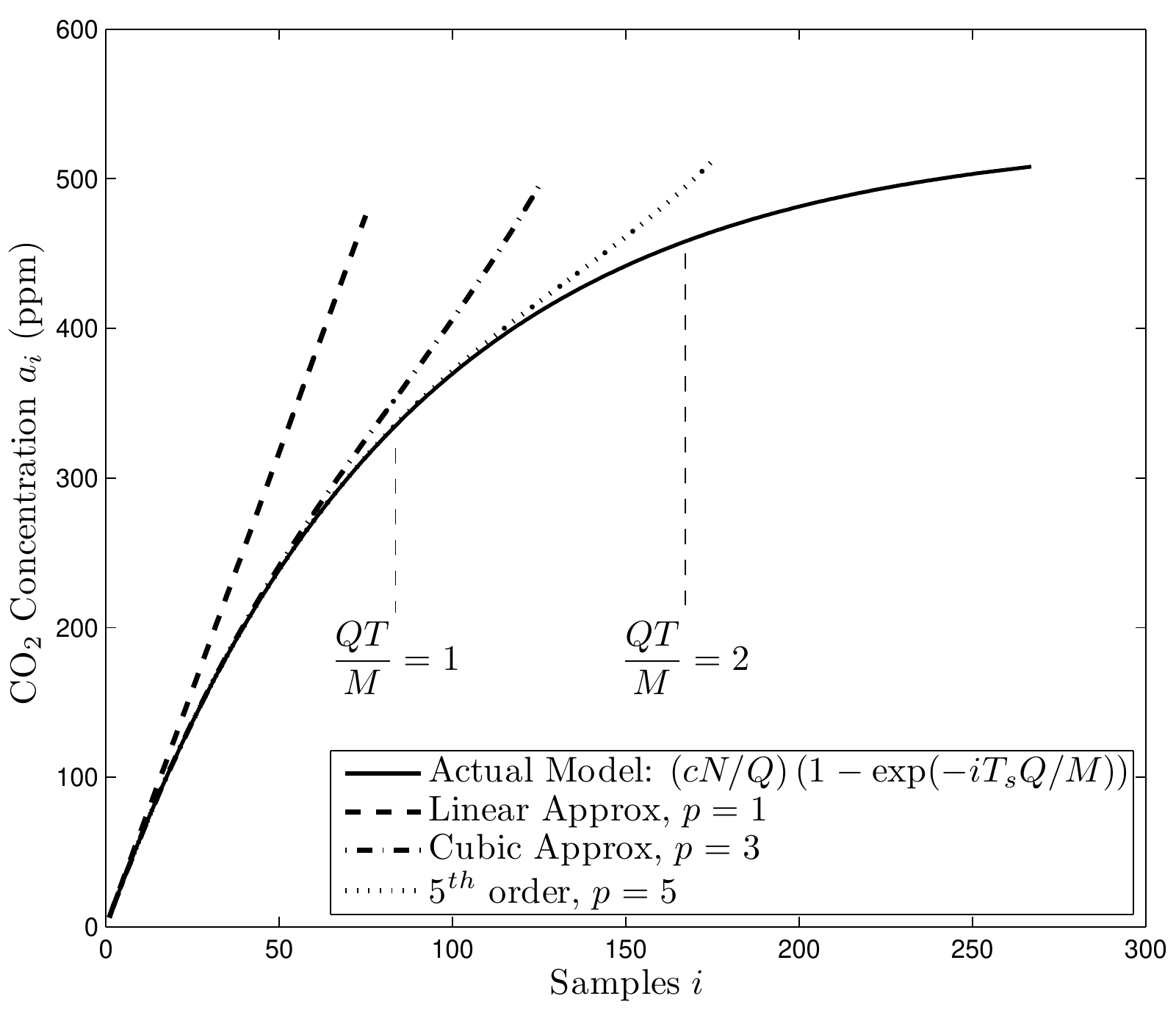}
  \caption{Growth curve predicted by the ideal indoor space model (solid line) and polynomial approximations of increasingly higher order (dashed lines). }
  \label{fig:model}
  \end{center}
\end{figure}

For the purpose of timely ventilation, the controller needs an estimate of $\widehat{N}$ as early as possible, preferably before equilibrium is reached. In other words, our desired operating region is the transient region of the curve. We note that the transient region can be approximated as a $p^{\text{th}}$ order polynomial,
\begin{align}
a_i&\approx (cN/Q)\left(1-\left(\sum_{j=0}^{p}\frac{(-iT_sQ/M)^j}{j!}  \right)\right).  \label{model:approx}
\end{align}
We define a parameter that is indicative of the transient region, $K$, as follows,
\begin{align}
K=\frac{QT}{M}, \label{defK}
\end{align}
which is the total amount of fresh air injected normalized with respect to the volume of the room. Figure \ref{fig:model} gives us an idea on the value of $p$ that describes $a_i$ with reasonable accuracy. We have considered the parameters corresponding to the experimental setup described later in \ref{sec:results:online}, i.e., one person in a $780$ cu.ft. room ventilated by fresh air at $28$ cu.ft./min. The resulting rise time is $M/(QT_s)=30$ samples and equilibrium value $cN/Q=530$ ppm.

\section{Related Work} \label{sec:relatedwork}
As discussed in Section \ref{sec:probform}, our goal is to estimate the occupancy $\widehat{N}$ and inflow-rate $\widehat{Q}$ from the observation sequence $\bo y=\bo a+\bo \epsilon$. In the non-linear regression literature (e.g., \cite{Kay93},\cite{Seb89}), the time series $\bo a$ given by \eqref{intmodel} is also known as Monomolecular Growth Curve. We discuss two existing approaches to solve this estimation problem and analytically evaluate their performance.

\subsection{Maximum Likelihood Estimation and CRLB} \label{sec:relatedwork:mle}
In the MLE approach \cite{Kay93}, we minimize the log-likelihood function $\mathcal{L}$ for the measurements $\bo y$, i.e.,
\begin{align}
\mathcal{L}(N,Q)&=\ln f(\bo{y};N,Q)\\
  &= -n\ln\left(\sqrt{2\pi}\sigma\right)- \frac{1}{2\sigma^2} \sum_{i=1}^n \left(y_i-a_i(N,Q) \right)^2,
\end{align}
and obtain the estimates $\widehat{N}$ and $\widehat{Q}$. Since $\epsilon_i$ are i.i.d. Gaussian, MLE is equivalent to minimizing the sum of squares
\begin{align}
R(N,Q) := \sum_{i=1}^n \left(y_i-a_i(N,Q)\right)^2.
\end{align}
The asymptotic variance of MLE attains the Cramer-Rao Lower Bound \cite{Kay93}, which is the inverse of the Fisher information matrix
\begin{align}
\mathcal{I}(N,Q)
&=\frac{1}{\sigma^2}\begin{pmatrix}
    \sum \left(\frac{\partial a_i}{\partial N}\right)^2   & \sum \frac{\partial a_i}{\partial N}\frac{\partial a_i}{\partial Q} \\
    \sum \frac{\partial a_i}{\partial N}\frac{\partial a_i}{\partial Q} & \sum \left(\frac{\partial a_i}{\partial N}\right)^2 \\
    \end{pmatrix}.
\end{align}
With the assumptions that (1) $T$ is chosen such that $K=QT/M$ is small (e.g., for our experimental setup, $QT/M\approx 0.5$ corresponding to $n=50$ samples) and (2) a second order $p=2$ is sufficient to describe the signal in \eqref{model:approx}, we obtain the following approximations for the partial derivatives
\begin{subequations}
\begin{eqnarray}
\partial a_i/\partial N \approx (cT_s/M) i, \mbox{ and } \\
\partial a_i/\partial Q \approx (cN/2)(T_s/M)^2 i^2.
\end{eqnarray}
\end{subequations}
With another assumption that (3) the number of measurements $n$ for $y_i$ is large, the CRLB can be derived as
\begin{align}
\text{CRLB} &\approx  \frac{48\sigma^2 T_s M^2}{T^3 c^2}.
\end{align}
When higher order polynomials are required to approximate the signal, using a similar approach, the CRLB can be more accurately expressed in the following form,
\begin{align}
\text{CRLB} \approx  \frac{48\sigma^2 T_s M^2}{T^3 c^2} \left(1+\alpha_1 K+\cdots+\alpha_j \frac{K^{j}}{j!} \right), \label{var:crlb:op}
\end{align}
where (1) $K=QT/M$, (2) higher $j$ implies more accuracy and (3) $\alpha_j$ are appropriate constants, e.g., the first three terms can be derived as $\alpha_1=2/3$, $\alpha_2=377/945$ and $\alpha_3=2411/11340$. We skip the derivations for the sake of brevity. The CRLB will be compared later with the performance metrics for other techniques in Section \ref{sec:results}.

\subsection{Recursive Least Squares based Estimation} \label{sec:relatedwork:rls}
The Kalman Filter based approach detailed in \cite{Fed97} simplifies to a Recursive Least Squares (RLS) algorithm (e.g., \cite{Kay93}) with the assumption of constant parameters. Based on the differential model for signal in \eqref{diffmodel}, the RLS model is predicated on the difference equation
\begin{align}
y_i-y_{i-1}=(T_s/M) (cN-Q a_i)+(\epsilon_i-\epsilon_{i-1}), \label{difference:eqn}
\end{align}
where $\epsilon_i$ represent the additive noise (observation error) and $y_i$ represents the measurements of the room concentration. Equation \eqref{difference:eqn} above represents a set of $n$ equations for the $n+1$ measurements for $i=0,1,\ldots,n$. Denoting $\eta_i$ as
\begin{align}
\eta_i=\epsilon_i-\epsilon_{i-1},
\end{align}
the equations can be represented as $\bo{Y}=\bo{X}\bo{\theta}+\bo{\eta}$, where $\bo{\theta}=[N,Q]'$ and $\bo{X}$ is the matrix containing the signal terms
\begin{align}
\bo{X}=\frac{T_s}{M}\begin{bmatrix}
          c & c & \cdots & c \\
          a_1 & a_2 & \cdots & a_n \\
        \end{bmatrix}'.
\end{align}
From the theory of Generalized Least Squares (e.g., \cite{Kay93}), we know that the LS estimate and the expected variance are given by
\begin{align}
\widehat{\bo{\theta}} &= \left[\bo{X'} \bo{V^{-1}} \bo{X}\right]^{-1}\bo{X'} \bo{V^{-1}} \bo{Y}, \mbox{ and } \\
\text{Var}(\widehat{\bo{\theta}}) &= \left[\bo{X'} \bo{V^{-1}} \bo{X}\right]^{-1},
\end{align}
where $\bo{V}=\text{Var}(\bo{\eta})$. We are interested in the variance of $\widehat{N}$, which is given by
\begin{align}
\text{Var}(\widehat{N})=\left[\left[\bo{X'}\bo{V^{-1}}\bo{X}\right]^{-1}\right]_{11}. \label{var:rls:approx}
\end{align}
The matrices $\bo{V}$ and $\bo{V^{-1}}$ need to be described further. Note that the time correlation of $\eta_i$ results in this $n\times n$ tridiagonal matrix
\begin{align}
\bo{V}&=\sigma^2\begin{bmatrix}
                               2 & -1 &   &   &   \\
                               -1 & 2 & -1 &   &   \\
                                 & \cdot & \cdot & \cdot &   \\
                                    &   & -1 & 2 & -1 \\
                                    &   &   & -1 & 2 \\
                             \end{bmatrix}.
\end{align}
The inverse of this tridiagonal matrix can be written in closed form. Bearing in mind that $\bo{V^{-1}}$ is symmetric, here we write only the upper diagonal terms
\begin{align}
\bo{V^{-1}}&=\frac{1}{\sigma^2(n+1)}\times \nonumber \\
&\begin{bmatrix}
                                      n & n-1 & \cdot & i & \cdot & 1 \\
                                        & 2(n-1) & \cdot & 2i & \cdot & 2 \\
                                        &   & \cdot & \cdot & \cdot & \cdot \\
                                        &   &   & (n-i+1)i & \cdot & n-i+1 \\
                                        \cdot&   &   &   & \cdot & \cdot \\
                                        1&\cdot   &   &   &   & n \\.
                                    \end{bmatrix}
\end{align}

Let $\sigma^2_{RLS}$ denote the variance given by \eqref{var:rls:approx}. Expressing the variance in a form similar to the CRLB \eqref{var:crlb:op} in Section \ref{sec:relatedwork:mle}, we obtain
\begin{align}
\sigma^2_{RLS} \approx  \frac{192\sigma^2 T_s M^2}{T^3 c^2} \left(1+\beta_1 K+\cdots+\beta_j \frac{K^{j}}{j!}\right), \label{var:rls:op}
\end{align}
where (1) $K=QT/M$, (2) higher $j$ implies more accuracy and (3) $\beta_j$ are appropriate constants, e.g., first three terms can be derived as $\beta_1=3/8$, $\beta_2=169/1120$ and $\beta_3=57/1120$.  We skip the derivations for the sake of brevity. This variance can be compared with the CRLB \eqref{var:crlb:op} at this point.

\section{Method of Moments based Estimation} \label{sec:mme}
In this paper, we adopt a Method of Moments (e.g., \cite{Kay93},\cite{New86}) based approach to estimate the occupancy and inflow-rate parameters. Unlike least squares approximation, the idea here is to match some theoretical and empirical properties of the model and solve the resulting equations to obtain the estimates.

For $m\geq 1$, let us define by $A_{m,T}(N,Q)$ the theoretical integral of $m^{th}$ power of $a_t(N,Q)$ in the time duration $[0,T]$,
\begin{align}
A_{m,T}(N,Q)&=\int_0^T \left( a_t(N,Q) \right)^m dt.
\end{align}
It is possible to obtain an empirical integral by approximating $\int_0^T y^m(t)dt$ with the help of the values $\{y_0,y_1,\ldots y_n\}$ at the set of time instants $\{0,T_s,2T_s,\ldots nT_s\}$. We define the approximation as $Y_{m,T}$,
\begin{align}
Y_{m,T}=\int_0^T y^m(t)dt,
\end{align}
so that the theoretical and empirical moments can be matched
\begin{align}
A_{m,T}(N,Q)&=Y_{m,T}.
\end{align}

Since we have to estimate two parameters, we can obtain two equations by matching any two moments. We must note here that the choice of moment functions is not unique and one has to base the choice on three considerations - (1) existence of a unique solution, (2) computational complexity in solving the equations and (3) expected variance of estimation error. In Section \ref{sec:mme:solmme}, we choose a family of moment functions that have a unique solution and the solution of which is amenable for recursive implementation as sample size increases. In Section \ref{sec:mme:var}, we find the expected variance of the estimator.

\subsection{Solution of Moment Equations} \label{sec:mme:solmme}
We consider the system of two equations matching the 1\sups{st} and any m\sups{th} ($m\geq 2$) moments. We use this to solve for our MME estimates $\widehat{N}$ and $\widehat{Q}$.
\begin{align}
A_{1,T}(N,Q)=Y_{1,T}, \mbox{ and } A_{1,T}(N,Q)=Y_{m,T}, \quad m\geq 2
\end{align}
From our model description in \eqref{intmodel}, $A_{m,T}(N,Q)$ can be described as
\begin{equation}
\begin{split}
A_{m,T}(N,Q)&= \left(\frac{cN}{Q}\right)^m T \int_0^1 \left(1-e^{-rz}\right)^m dz, \\
 \mbox{ where } r&=QT/M, z=t/T.
\end{split} \label{defA}
\end{equation}
From the expressions for $A_{m,T}$ in \eqref{defA}, one can observe that these are non-linear equations in $N$ and $Q$. We demonstrate that they have a unique solution by presenting an approach in which these solutions can be obtained. We define a function $G_m(r)$, obtained by eliminating $N$ from the expressions for $A_{1,T}$ and $A_{m,T}$, as follows
\begin{align}
G_m(r) &= \frac{T^{m-1} A_{m,T}}{\left(A_{1,T}\right)^m} \nonumber\\
       &= \frac{\int_0^1 \left(1-e^{-rz}\right)^m dz}{\left(\int_0^1 \left(1-e^{-rz}\right) dz\right)^m} . \label{defG}
\end{align}
It is easy to show that $G_m(r)$ is a decreasing convex function in $(0,\infty)$ for $m=2$ and this property appears to be true for all $G_m(r)$ for $m\geq 2$. Monotonicity of $G_m(r)$ ensures the existence of an inverse while convexity of $G_m(r)$ ensures that the inverse can be computed quickly using any gradient-based numerical method. The solution for inflow-rate is
\begin{align}
\widehat{Q} = \frac{M}{nT_s}G_m^{-1}\left(\frac{T^{m-1} Y_{m,T}}{\left(Y_{1,T}\right)^m} \right).
\end{align}
We use the facts (1) $Y_{1,T}=A_{1,T}(N,Q)$ (the first moment condition) and (2) $A_{1,T}(N,Q)=N A_{1,T}(1,Q)$ (follows from \eqref{defA}), to obtain the estimate of the number of occupants
\begin{align}
\widehat{N} = \frac{M}{cT_s} \frac{Y_{1,T}}{A_{1,T}(1,\widehat{Q})}.
\end{align}
The existence of a unique solution means that our MME estimator is well defined. From the theory of Method of Moments \cite{New86}, this uniqueness property also implies statistical consistency of the estimators, i.e., given a large number of samples, the estimates converge to the true values.

\subsection{Expected variance of occupancy} \label{sec:mme:var}
To compare the performance of the MME method with that of MLE or RLS, we derive the expected variance of our MME estimator for the number of occupants $\widehat{N}$. First, let us define the following vectors
\begin{align}
\quad \bo{A}=(A_{1,T},A_{m,T})',\mbox{ and } \quad \bo{Y}=(Y_{1,T},Y_{m,T})'
\end{align}
so that our MME estimates satisfy
\begin{align}
\bo{A}(\widehat{\bo{\theta}})= \bo{Y}, \label{Ath:is:Y}
\end{align}
where $\bo{\theta}=[N,Q]'$, For small estimation errors (i.e., small values of $\sigma^2$), one can obtain the Taylor series approximation of $\bo{A}(\widehat{\bo{\theta}})$ around $\bo{\theta}$
\begin{align}
\bo{A}(\widehat{\bo{\theta}}) \approx \bo{A}\left(\bo{\theta}\right)+ \frac{\partial\bo{A(\bo{\theta})}}{\partial\bo{\theta}'} \left(\widehat{\bo{\theta}}-\bo{\theta}\right).
\end{align}
This can be rewritten in the following form
\begin{align}
\widehat{\bo{\theta}}-\bo{\theta} \approx \left[\frac{\partial\bo{A(\bo{\theta})}}{\partial\bo{\theta}'} \right]^{-1} \left(\bo{A}(\widehat{\bo{\theta}}) -\bo{A}(\bo{\theta})\right). \label{Theta:taylor}
\end{align}
We denote
\begin{align}
\bo{H}= \left[\frac{\partial\bo{A(\bo{\theta})}}{\partial\bo{\theta}'}\right]^{-1}, \mbox{ and } \bo{\Sigma}=\text{Var}\left(\bo{A}(\widehat{\bo{\theta}})\right). \label{defKS}
\end{align}
Asymptotic normality of $\bo{A}(\widehat{\bo{\theta}})$, which we show later, and \eqref{Theta:taylor} then implies $\text{Var}(\widehat{\bo{\theta}}) \approx \bo{H}\bo{\Sigma}\bo{H'}$. We are interested in $\text{Var}(\widehat{N})$, which is actually $\left[\text{Var} (\widehat{\bo{\theta}}) \right]_{11}$, and hence
\begin{align}
\text{Var}(\widehat{N}) \approx \left[\bo{H}\bo{\Sigma}\bo{H'}\right]_{11}. \label{var:mme:approx}
\end{align}
Next, we derive simplified expressions for $\bo\Sigma$ and $\bo H$.

We note from \eqref{Ath:is:Y} and \eqref{defKS} that $\bo{\Sigma}=\text{Var} \bo Y$. If the sampling interval is small enough compared to the rate of change of $y_t$, we can use constant interpolants $Y_{m,T}=T_s \sum y^m_i$. Therefore, $\text{Var}\left(Y_{m,T} \right) = T_s^2 \sum \text{Var}\left(y^m_i\right)$. Hence, for $m=1$,
\begin{align}
\text{Var}\left(Y_{1,n}\right) = T_s^2  \sum \text{Var}\left(y_i\right) =  T_s^2 n\sigma^2.
\end{align}
For calculating the variance for $m\ge 2$, we consider the expansion of $y^m_i=(a_i+\epsilon_i)^m$ and write their expectation as follows
\begin{align}
\begin{split}
y^m_i&= a^m_i+m a^{m-1}_i \epsilon_i + \binom{m}{2} a^{m-2}_i \epsilon_i^2 +\cdots \\
\mathbb{E}\left(y^m_i\right)&=a^m_i + \binom{m}{2} a^{m-2}_i \sigma^2 + \cdots
\end{split} \label{ymi:expand}
\end{align}
The influence of the noise terms $\epsilon_i,\epsilon_i^2,\ldots$ on the overall variance decays exponentially since they are multiplied by decaying factors $a^{m-1}_i,a^{m-2}_i$, etc. Hence, we can approximate $y^m_i$ by retaining only the first two terms from \eqref{ymi:expand}, which means that $y^m_i$ is approximately Gaussian with mean $a^m_i$ and variance
\begin{equation}
\begin{split}
\text{Var}\left(y^m_i\right)&= \text{Var}\left(y^m_i - \mathbb{E}\left(y^m_i\right)\right) \\
&= \text{Var} \left(m a^{m-1}_i \epsilon_i+ O\left(a^{m-2}_i\right)\left(\epsilon_i^2-\sigma^2\right)+\cdots \right)\\
&\approx \text{Var} \left(m a^{m-1}_i \epsilon_i\right).
\end{split}
\end{equation}
Next, we can complete our derivation
\begin{align}
\text{Var}\left(Y_{m,T}\right) &= T_s^2 \sum \text{Var}\left(m a^{m-1}_i \epsilon_i\right) \nonumber \\
&= T_s^2  m^2\sigma^2 \sum a^{2m-2}_i
\end{align}
Using exactly similar arguments, the covariance term is
\begin{align}
\text{Cov}\left(Y_{1,T},Y_{m,T}\right) &\approx T_s^2 m \sigma^2 \sum a^{m-1}_i,
\end{align}
so that $\bo{A}(\widehat{\bo{\theta}})$ is approximately normal with covariance matrix
\begin{align}
\bo{\Sigma}\approx T_s^2 \sigma^2 \left[
                                \begin{array}{cc}
                                  n & (cNT_s/M)^{m-1}n^m \\
                                  (cNT_s/M)^{m-1}n^m & m^2 (cNT_s/M)^{2m-2} \frac{n^{2m-1}}{2m-1} \\
                                \end{array}
                              \right],   \label{Sigma:cov}
\end{align}
with the assumptions that (1) the signal in \eqref{model:approx} is adequately described by $p=2$ and (2) number of observations $n$ is large.

Assuming small $QT/M$, we obtain the following first order approximation of $A_{m,T}$
\begin{align}
A_{m,T} &\approx T_s \left(\frac{cNT_s}{M} \right)^m \left(\frac{n^{m+1}}{m+1} - \frac{m QT_s}{2M}\frac{n^{m+2}}{m+2}\right),
\end{align}
from which we can obtain $\bo{H}$. Using the expressions for $\bo{H}$ and $\bo{\Sigma}$ in \eqref{var:mme:approx}, we obtain
\begin{align}
\text{Var}(\widehat{N}) &\approx \left(\frac{M}{cT_s}\right)^2\frac{4(m+1)^2(12+m)}{m(2m-1)}\frac{\sigma^2}{n^3}. \label{var:mme:m}
\end{align}
For an MME algorithm with moment $m=2$, we denote the variance as $\sigma^2_{MME,2}$, so that
\begin{align}
\sigma^2_{MME,2} \approx \frac{84\sigma^2 T_s M^2}{T^3 c^2}.
\end{align}
Relaxing the assumption of small $QT/M$, we can express the variance in a form similar to the CRLB \eqref{var:crlb:op} in Section \ref{sec:relatedwork:mle}.
\begin{align}
\sigma^2_{MME,2} \approx  \frac{84\sigma^2 T_s M^2}{T^3 c^2} \left(1+\gamma_1 K+\cdots+\gamma_j \frac{K^{j}}{j!} \right), \label{var:mme:op}
\end{align}
where (1) $K=QT/M$, (2) higher $j$ implies more accuracy and (2) $\gamma_j$ are appropriate constants, e.g., first three terms can be derived as $\gamma_1=24/35$, $\gamma_2=11/25$ and $\gamma_3=1581/6125$.  We skip the derivations for the sake of brevity. The variance can now be compared to the CRLB (given by \eqref{var:crlb:op}) and RLS (given by \eqref{var:rls:op}). We can see that for small $QT/M$, expected variances of all of these techniques are proportional to the variance of the observation noise and inversely proportional to the cube of the number of samples. However, they have different multiplicative factors. The MLE has the best performance (factor of 48), followed by MME (factor of 84) and RLS (factor of 192).

\subsection{Notes on Implementation} \label{sec:mme:notes}
\begin{figure}[htb]
\begin{center}
    \includegraphics[width=9cm]{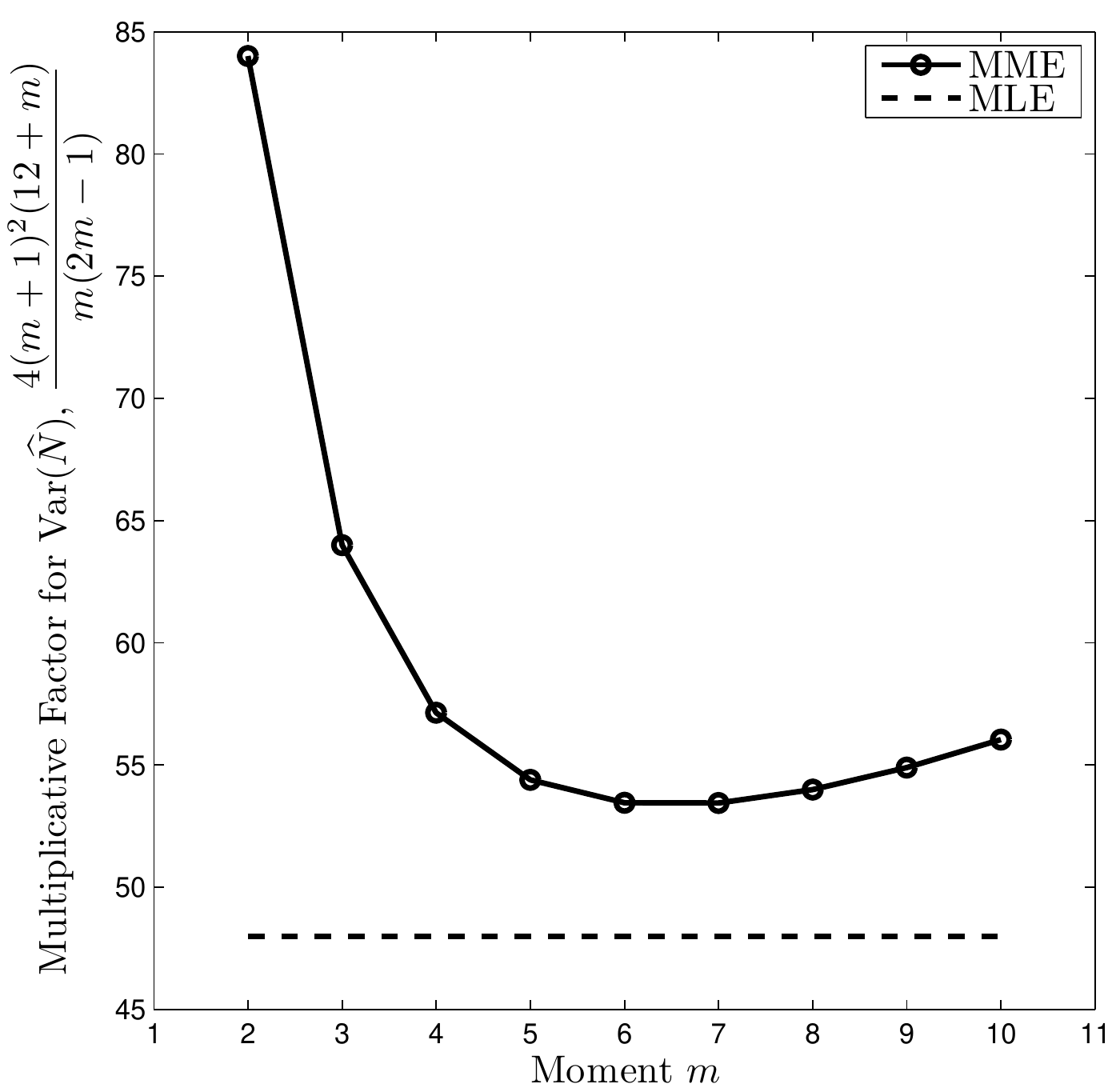}
  \caption{Performance of MME based estimation in the transient region for different chosen moments.}
  \label{fig:mmevar}
  \end{center}
\end{figure}

Some more remarks can be made about the performance of the MME estimator, given by \eqref{var:mme:m}, for a general value of $m$. For various moments $m$ and small $QT/M$, we plot the multiplicative factors associated with $\sigma^2 T_s M^2/(T^3 c^2)$ in Figure \ref{fig:mmevar}.

Choosing higher order moments decreases the variance initially before it starts increasing again. For example, for $m=3$, the multiplicative factor is reduced to $64$. The minima is attained for $m\approx 6$ when the factor reduces to $53.4$. Since MME is a sub-optimal method, its variance is always greater than CRLB which has a multiplicative factor of only $48$. It may be noted that estimation using higher moments involves more computation and hence we have considered only $m=2,3$ in this paper for purposes of illustration.

We remark on the computational efficiency of the estimator. Two components of our procedure require numerical computations - the integration of $y_t$ to obtain $Y_{m,T}$ and the inverse of the function $G_m(r)$ to obtain $\widehat{r}$. Regarding the numerical integration, the trapezoidal method (e.g., \cite{Kin02}) is readily amenable for recursive implementation. Regarding the inverse, monotonicity and convexity of $G_m(r)$ (which can be established easily and skipped here for brevity) ensure that any iterative gradient based method can be used to compute the inverse. We have used Newton's Method in our simulations.

\section{Illustrative Examples} \label{sec:results}

\subsection{Experimental setup and online estimation} \label{sec:results:online}
Experiments were performed in the Building Energy and Environmental Systems Laboratory (BEESL) at Syracuse University. We used a chamber of dimension $6.5$ ft by $12$ ft by $10$ ft high that was connected to an HVAC system with an independent roof top unit. The fresh air supply was set at a constant rate of $Q=28$ CFM (cubic feet per minute). We use $Q$ to know the ground truth but treat it as an unknown in our estimation procedure. To ensure well mixed conditions, a fan was used inside the chamber. \co concentration inside the room was measured (in parts per million volume) using Graywolf sensor IQ-610, with a rated accuracy of $\pm 3\%$ reading and $\pm 50$ ppm. Constant inflow of \co gas was injected into the chamber using Alicat Scientific mass flow controller with a rated accuracy of $\pm 0.8\%$ reading and $\pm 0.2\%$ full scale, where full scale in this case was $20$ SLPM (standard litres per minute). The source was a commercial gas cylinder containing $100\%$ \co. The generation rate was maintained at $0.42$ SLPM or $1.4 \times 10^{-5}$ kg/sec, which corresponds to the rate at which a $170$ pound person would produce \co while standing and performing a light task such as drawing \cite{McA91}. This rate was also used as a benchmark in \cite{Fed97}. The sampling interval $T_s$ was chosen as $20$ sec. The time constant of the growth curve can be estimated from these settings as $M/Q\approx 83$ samples or $28$ minutes. Injection of gas source into the chamber was started only $2$ hours after closing the door, so as to ensure completely fresh air inside the chamber. The fresh air concentration $C_0$ was taken as the mean of $30$ samples before the source was injected and was measured to be $392$ ppm.

\begin{figure}[!h]
\begin{center}
    \includegraphics[width=9cm]{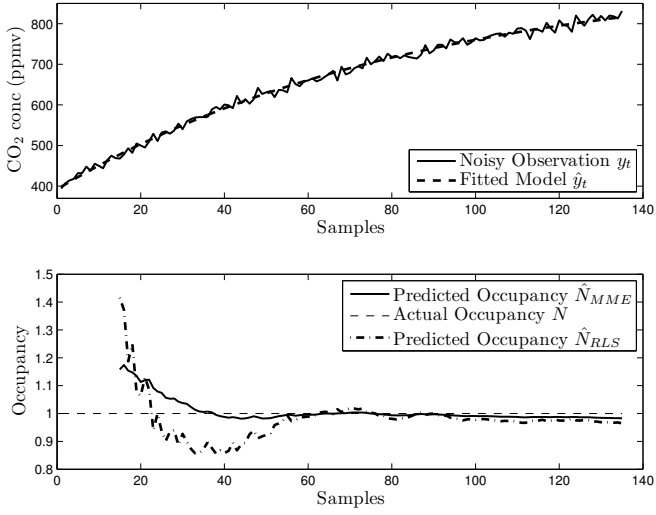}
  \caption{Online occupancy ($\widehat{N}$) estimation using experimental data, using MME and RLS techniques.}
  \label{fig:tracking}
  \end{center}
\end{figure}

A sample online performance of the MME and RLS based estimators is shown in Figure \ref{fig:tracking}. Since $N=1$, the occupancy estimator converges $\widehat{N}\rightarrow 1$ after sufficiently large number of samples.
The top part of Figure \ref{fig:tracking} shows the actual noisy observations. The steady state \co level can be computed as $C_0+cN/Q\approx 922$ ppm. As we can see, a reasonably accurate estimate can be obtained using fairly small number of samples, i.e., much before equilibrium is reached in the observed concentration. Also, the MME method provides estimates that appear less oscillatory and have lower variance compared to the RLS counterpart. We compare the statistical performances more rigorously in the next subsection by using Monte-Carlo simulations.

\subsection{Performance comparison using Monte-Carlo simulations} \label{sec:results:monte}
\begin{figure}[!h]
\begin{center}
    \includegraphics[width=9cm]{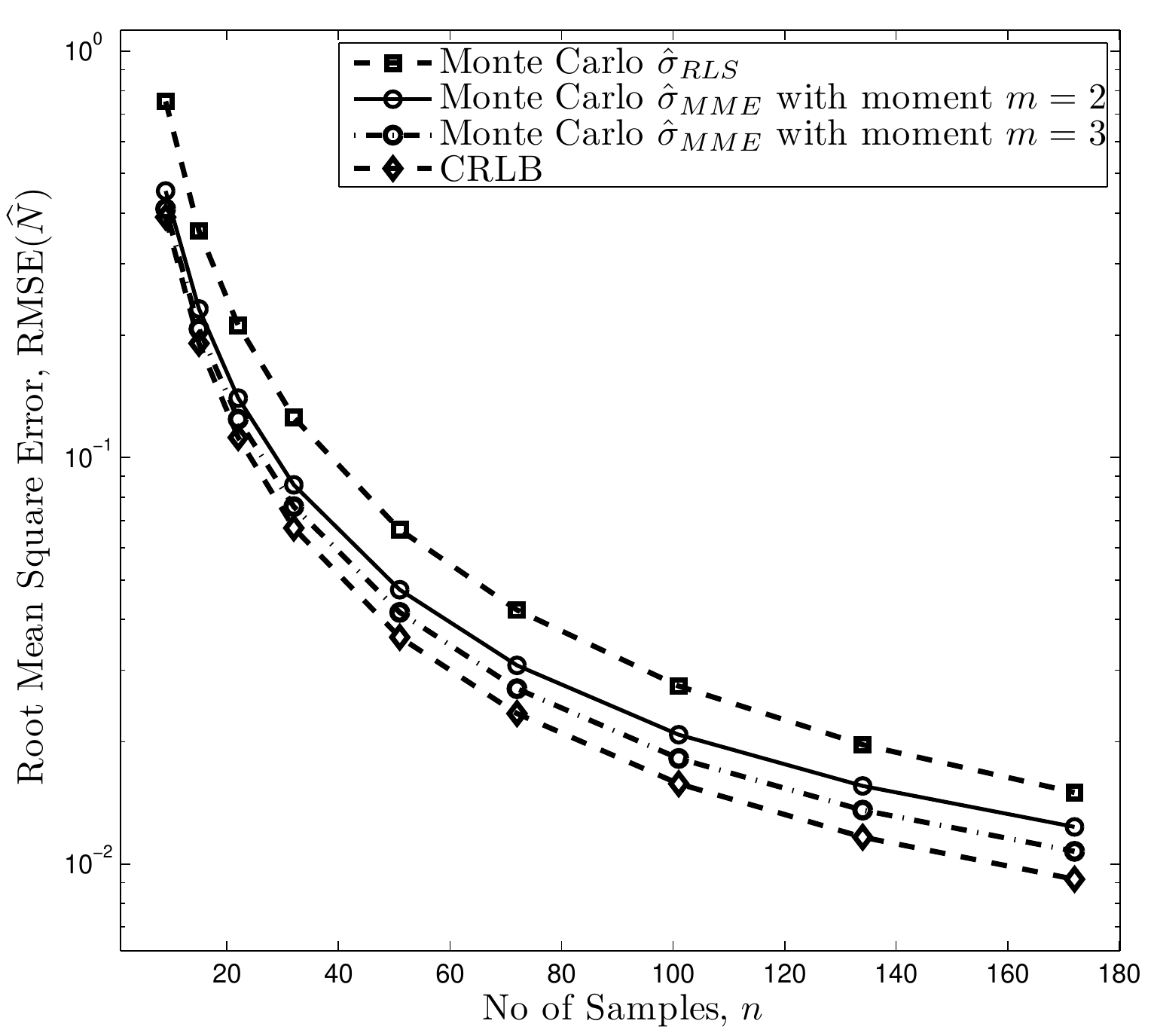}
  \caption{Performance comparison of MME, RLS and MLE schemes using Monte-Carlo simulations, assuming $\sigma=10$ ppm and $T_s=20$ sec.}
  \label{fig:monte}
  \end{center}
\end{figure}

In our experiments with the IQ-610 sensor, the steady state standard deviation was found to be approximately $8$ ppm. Hence for our simulations, we assume the measurement standard deviation $\sigma = 10$ ppm. To compare the performance of various schemes, we use the root mean square error of the occupancy estimate $\sqrt{\text{Var}(\widehat{N})}$. We compare 4 schemes, namely MME with moments of order 2 and 3, RLS and MLE. The simulation setup is similar to that in Section \ref{sec:results:online}. The results for a varying number of samples are shown in Figure \ref{fig:monte}. We observe that an MME scheme with the moment $m=3$ performs better than the one with the moment $m=2$. Both the MME schemes perform better than the RLS procedure. The relative performances of these schemes are partly predicted from the estimation variances derived in \eqref{var:crlb:op},\eqref{var:rls:op} and \eqref{var:mme:op}, and also the discussion in Section \ref{sec:mme:notes}. For quick reference, those metrics can be recalled as $\text{CRLB}\propto 48\sigma^2/T^3$, $\sigma^2_{MME,3}\propto  64\sigma^2/T^3$, $\sigma^2_{MME,2}\propto  84\sigma^2/T^3$ and $\sigma^2_{RLS}\propto  192\sigma^2/T^3$.

\begin{figure}[!h]
\begin{center}
    \includegraphics[width=9cm]{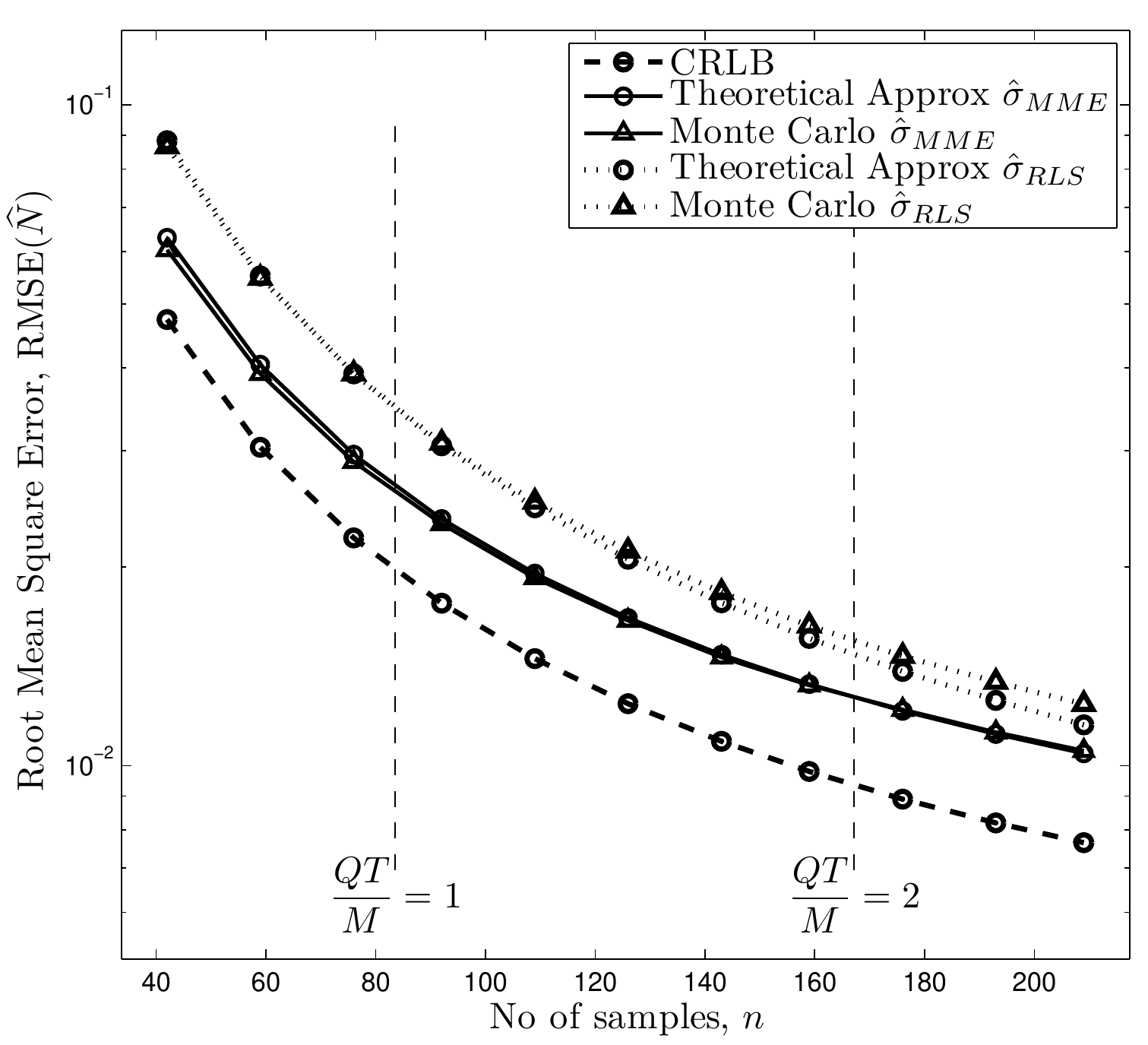}
  \caption{Theoretical approximations compared with Monte-Carlo performance, assuming $\sigma=10$ ppm and $0.4\le QT/M \le 2.5$.}
  \label{fig:Ts10sig5}
  \end{center}
\end{figure}

Next, we evaluate the accuracy of the transient-region expressions for variance derived in this paper. We use the third order expressions for the RLS (given by \eqref{var:rls:op}) and MME (given by \eqref{var:mme:op}) techniques. The constants $\alpha_j,\beta_j$ and $\gamma_j$ for $j\le 3$ were given earlier in Sections \ref{sec:relatedwork:mle}, \ref{sec:relatedwork:rls} and \ref{sec:mme:var} respectively. We display both the theoretical performance predictions and Monte Carlo results in Figure \ref{fig:Ts10sig5}, which are found to be extremely close.

\subsection{Different metabolic rates} \label{sec:results:mulN}
\begin{figure}[!h]
\begin{center}
    \includegraphics[width=9cm]{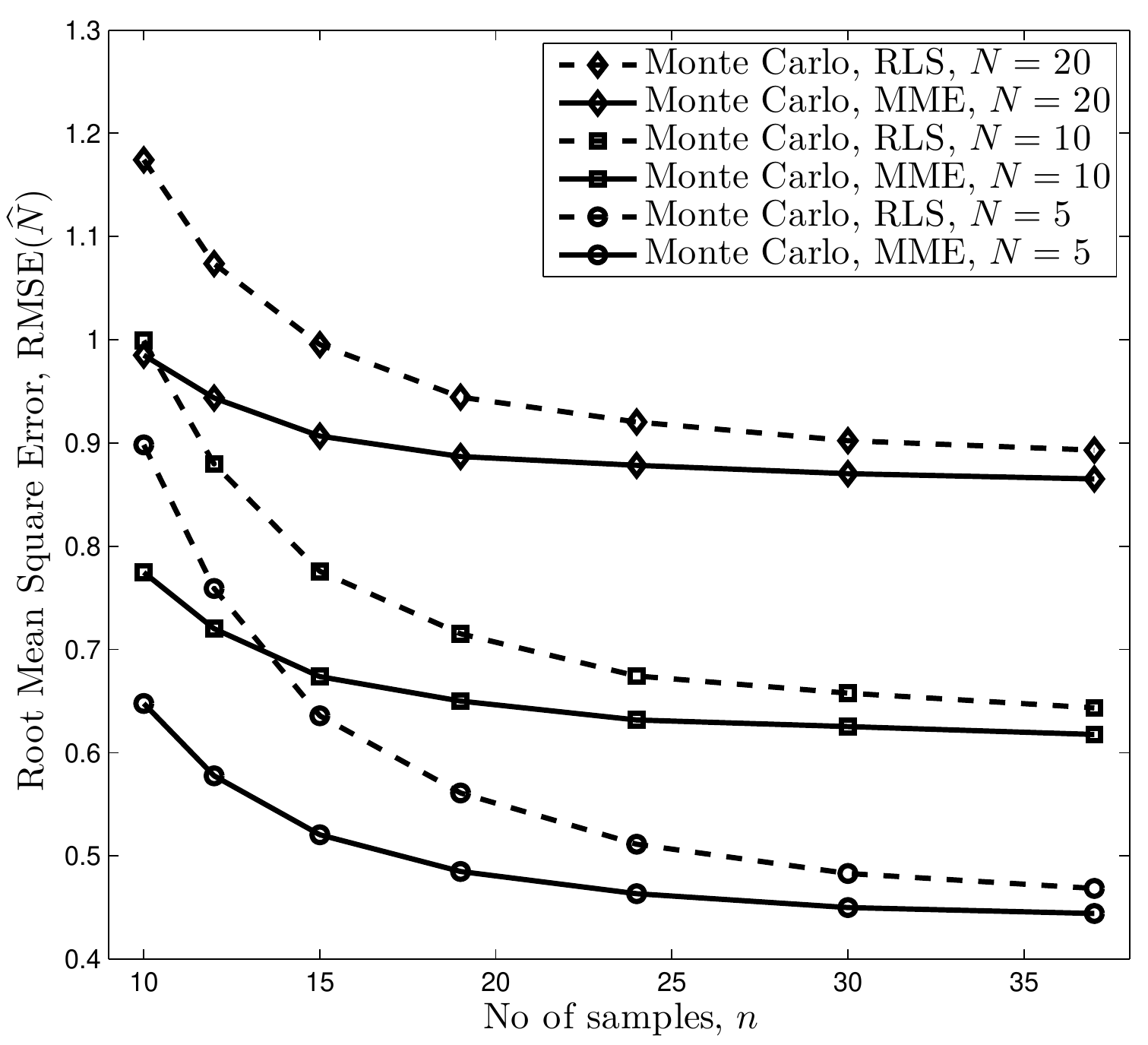}
  \caption{Estimation error due to multiple persons with different metabolic rates.}
  \label{fig:mulN}
  \end{center}
\end{figure}

An additional source of error in estimating occupancy from the \co generation rate is the fact that the volumetric rate of \co generation, $c$, depends on many factors such as Dubois body surface area $A_D$, the metabolic heat produced by the body $M_H$ and the respiration quotient $RQ$, which is the ratio of \co exhaled to oxygen inhaled. Based on ASHRAE (e.g., \cite{Aglan03},\cite{Persily97}), the volumetric rate of \co generation per person can be written as
\begin{equation}
c=\frac{0.0028 A_DM_H R_Q}{0.23 R_Q+0.77}.
\end{equation}
For an average size man $A_D=1.8$ mt\sups{2}. Metabolic rate $M_H$ depends strongly on the various types of activities and can range between 1 and 2 for the occupants of an office building \cite{Persily97}. For an adult of average size engaged in light sedentary activity $R_Q$ is about $0.83$. For our simulations, we have replicated an experimental classroom setup in \cite{Aglan03}, where the volume of the room is $M=6143$ cu.ft., and the supply inflow-rate $Q=115$ cfm. The sampling period was taken as $T_s=2.5$ min.  We have considered population sizes of $5$, $10$ and $20$ and for all the cases we assume that each person generates \co with a metabolic rate uniformly distributed between 1 and 2. While estimating $N$, we assume that all the persons have a metabolic rate of $M=1.5$. The results from $10^5$ Monte Carlo trials are summarized in Figure \ref{fig:mulN}. Because of the uncertainty in generation rates for each person, the estimates for the number of persons are inconsistent, i.e., do not converge to zero even for a large number of samples. The MME method is found to have better prediction capability for all cases.

\subsection{Changing Occupancy} \label{sec:results:rwkN}
\begin{figure}[!h]
\begin{center}
    \includegraphics[width=9cm]{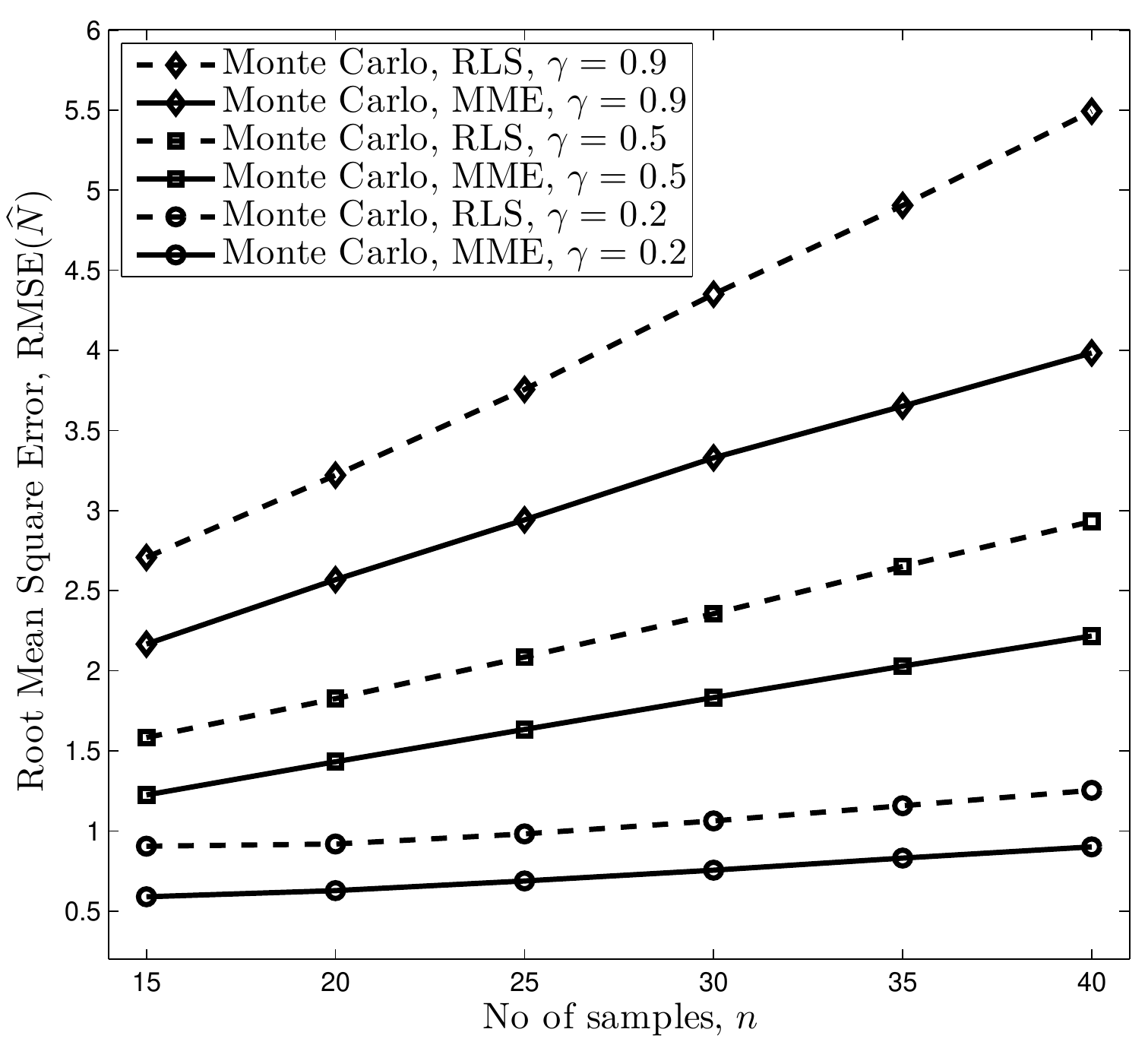}
  \caption{Estimation error due to changing occupancy.}
  \label{fig:rwkN}
  \end{center}
\end{figure}

One of our assumptions for both the RLS and MME estimation procedures has been that the occupancy is constant. In this subsection, we consider the case when the occupancy changes slightly during the duration of estimation. This is indicative of the situation when some attendees join or leave the meeting/ classroom. We consider a random walk model for occupancy, which was also used in \cite{Fed97}. We denote the continuous occupancy state as $N'$, which is modeled as a random walk
\begin{equation}
N'_{i+1}=N'_i+\kappa_i, \mbox{ with } \kappa_i\stackrel{i.i.d.}{\sim}\mathcal{N}(0,\gamma^2),
\end{equation}
where $\kappa_i$ are the disturbances that represent changing occupancy and $\gamma^2$ is the variance. The actual occupancy is modeled by the integer equivalent of $N'$, i.e.,
\begin{equation}
N_i= \lfloor N'_i \rfloor.
\end{equation}
The carbon dioxide concentration is then calculated using the integral form of the mass balance equation \eqref{diffmodel} for each time step. The initial number of occupants is assumed to be $N_0=20$ persons. Rest of the simulation setup is similar to that in Section \ref{sec:results:mulN}, i.e., $M=6143$ cu.ft., $Q=115$ cfm, $T_s=2.5$ min. The estimation error is measured relative to the mean of the occupancy for the entire period and the results are displayed in Figure \ref{fig:rwkN} for three different variances $\gamma=0.2,0.5$ and $0.9$. Since a random walk has a linearly increasing variance, the estimation error in Figure \ref{fig:rwkN} increases with time. Also, the results show that the MME technique has better estimation capability than the RLS procedure.

\subsection{Timing and Sensor Reductions} \label{sec:results:metrics}
The improvement of estimation accuracy has direct implications for other important system parameters, e.g., reduction in time to estimate and in number of sensors. From the transient region expressions of the variance, for small $QT/M$, we have for RLS \eqref{var:rls:op} and MME \eqref{var:mme:op},
\begin{equation}
\sigma^2_{RLS}\propto 192 \sigma^2/T^3, \mbox{ and } \sigma^2_{MME,2}\propto 84 \sigma^2/T^3,
\end{equation}
with identical constants of proportionality. In other words, for the same levels of acceptable variance, the RLS procedure would require $\sqrt[3]{192/84}\approx 1.3$ times the number of samples, i.e., $30\%$ more than the MME procedure. To the ventilation system, this will mean significant reduction in delay to respond.

Also, it is well known that readings from multiple sensors can be fused to improve the overall accuracy \cite{Kay93}. In particular, $N_s$ independent sensors are known to reduce the variance of estimates by $1/N_s$. If we consider a room instrumented with multiple and redundant sensors, then the RLS procedure would require $192/84\approx 2.28$ times more number of sensors than the MLE technique. This may translate to significant savings in terms of instrumentation cost.

\section{Conclusion} \label{sec:conclude}
In this paper, we have proposed a new approach for solving the problem of estimating the strength of a gaseous source in a room. We have assumed that the room is well mixed and a simplified mass balance equation is sufficient to describe the dynamics of the concentration over time. Also, it was assumed that the noise in the measurements is additive and Gaussian in nature. Since we are interested in making estimations as quickly as possible and much before equilibrium is reached, we have selected the initial section of the model dynamics as our operating region. We have performed a theoretical analysis of the estimation performances for our technique as well as another existing technique, and also obtained simplified expressions for this operating region. We have compared the performances of the two techniques using Monte-Carlo simulations and compared both with the lower bound. Our results clearly indicate the superiority of our approach. The proposed algorithm can potentially improve the performance of building control systems, since it can provide fast and accurate estimates of the strength of a gaseous source inside an indoor environment.

\section*{Acknowledgment}
The authors gratefully acknowledge support for this work by Syracuse Center of Excellence CARTI project award, which is supported by a grant from U.S. Environmental Protection Agency [Award No: X-83232501-0]. The authors would like to thank Dr. H. Ezzat Khalifa, Dr. Jianshun Zhang, Jingjing Pei and James Smith for their valuable advice and help with the experiments. The authors would also like to thank the anonymous reviewers for their valuable comments which helped us improve the paper substantially.

\bibliographystyle{IEEEtran}
\bibliography{mybib}

\begin{IEEEbiography} 
[{\includegraphics[width=1in,height=1.25in,clip,keepaspectratio]{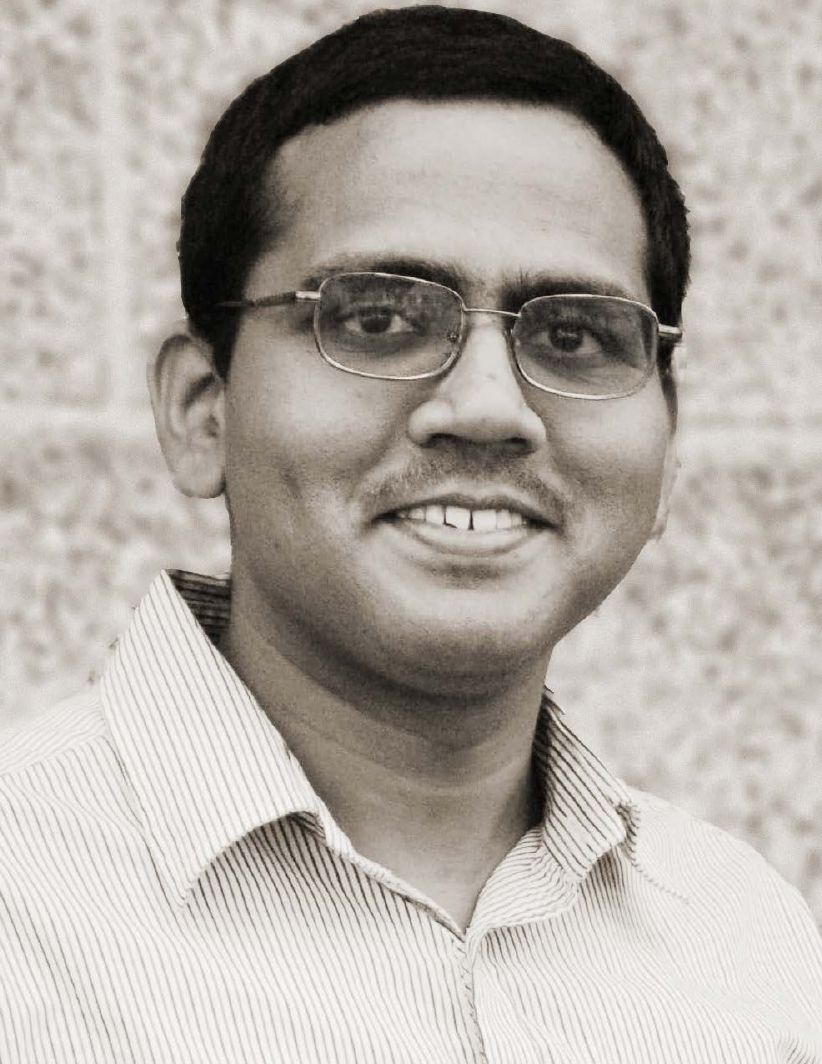}}]{Swarnendu Kar}
(S'07) received the B.Tech. degree in electronics and electrical communication engineering from the Indian Institute of Technology, Kharagpur, India, in 2004 and the M.S. degree in mathematics from Syracuse University, Syracuse, NY, in 2009, where he is currently working toward the Ph.D. degree in electrical engineering.
He was a Video Systems Engineer with Ittiam Systems Pvt. Ltd., Bangalore, India, during 2004--2006. He was a visiting student at The University of Melbourne, Parkville, Australia, during October--December 2009. His research interests include detection and estimation theory and distributed estimation in the context of sensor networks.
\end{IEEEbiography}

\begin{IEEEbiography} 
[{\includegraphics[width=1in,height=1.25in,clip,keepaspectratio]{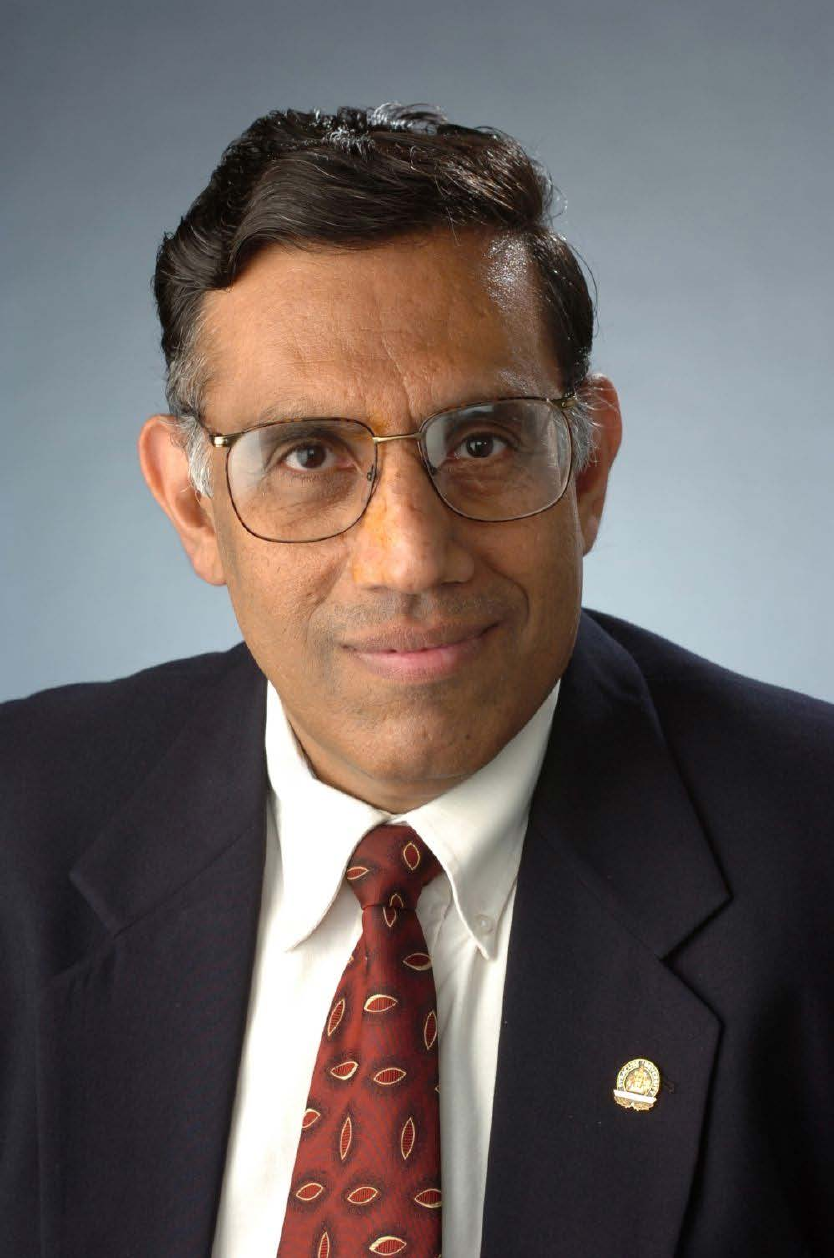}}]{Pramod K. Varshney}
(S'72-–M'77-–SM'82-–F'97) was born in Allahabad, India, on July 1, 1952. He received the B.S. degree in electrical engineering and computer science (with highest hons.), and the M.S. and Ph.D. degrees in electrical engineering from the University of Illinois at Urbana-Champaign in 1972, 1974, and 1976 respectively. From 1972 to 1976, he held teaching and research assistantships at the University of Illinois. Since 1976, he has been with Syracuse University, Syracuse, NY, where he is currently a Distinguished Professor of Electrical Engineering and Computer Science and the Director of CASE: Center for Advanced Systems and Engineering. He served as the Associate Chair of the department from 1993 to 1996. He is also an Adjunct Professor of Radiology at Upstate Medical University, Syracuse, NY. His current research interests are in distributed sensor networks and data fusion, detection and estimation theory, wireless communications, image processing, radar signal processing, and remote sensing. He has published extensively. He is the author of \emph{Distributed Detection and Data Fusion} (Springer-Verlag, 1997). He has served as a consultant to several major companies. Dr.Varshney was a James Scholar, a Bronze Tablet Senior, and a Fellow while at the University of Illinois. He is a member of Tau Beta Pi and is the recipient of the 1981 ASEE Dow Outstanding Young Faculty Award. He was elected to the grade of Fellow of the IEEE in 1997 for his contributions in the area of distributed detection and data fusion. He was the Guest Editor of the Special Issue on Data Fusion of the PROCEEDINGS OF THE IEEE, January 1997. In 2000, he received the Third Millennium Medal from the IEEE and Chancellor’s Citation for exceptional academic achievement at Syracuse University. He is the recipient of the IEEE 2012 Judith A. Resnik Award. He serves as a Distinguished Lecturer for the IEEE Aerospace and Electronic Systems (AES) Society. He is on the Editorial Board of the \emph{Journal on Advances in Information Fusion}. He was the President of International Society of Information Fusion during 2001.
\end{IEEEbiography}

\end{document}